\documentclass[10pt,letterpaper]{article}
\usepackage{express}

\begin{document}

\title{One-step implementation of a hybrid Fredkin gate with quantum memories and single superconducting qubit in circuit QED and its applications}
\author{Tong Liu, Bao-Qing Guo, Chang-Shui Yu,* and Wei-Ning Zhang}
\address{School of Physics, Dalian University of Technology, Dalian 116024, China}
\email{*ycs@dlut.edu.cn}
\date{\today}

\begin{abstract}
In a recent remarkable experiment [R. B. Patel \textit{et al.}, Science
advances 2, e1501531 (2016)], a 3-qubit quantum Fredkin (i.e.,
controlled-SWAP) gate was demonstrated by using linear
optics. Here we propose a simple experimental scheme by utilizing the
dispersive interaction in superconducting quantum circuit to implement a
hybrid Fredkin gate with a superconducting flux qubit as the control qubit
and two separated quantum memories as the target qudits. The quantum
memories considered here are prepared by the superconducting coplanar
waveguide resonators or nitrogen-vacancy center ensembles. In particular, it
is shown that this Fredkin gate can be realized using a single-step
operation and more importantly, each target qudit can be in an arbitrary state
with arbitrary degrees of freedom. Furthermore, we show that this
experimental scheme has many potential applications in quantum computation and quantum
information processing such as generating arbitrary entangled states
(discrete-variable states or continuous-variable states) of the two
memories, measuring the fidelity and the entanglement between the two
memories. With state-of-the-art circuit QED technology, the numerical simulation
is performed to demonstrate that two-memory NOON states, entangled coherent
states, and entangled cat states can be efficiently synthesized.
\end{abstract}






\section{Introduction}

\label{sec1}
The Fredkin gate is a three-qubit controlled-SWAP gate. Conditioned on the state
of the control qubit, the gate can enable the two target qubits swap their quantum states \cite%
{Fredkin}. It has played an important role in quantum computation and
quantum information processing (QCQIP) such as error correction \cite%
{correction1,correction2}, quantum cloning \cite{cloning}, quantum
fingerprinting \cite{fingerprinting1,fingerprinting2}, and quantum digital
signatures \cite{digital}. Any multiqubit gate can in principle be
decomposed into a sequence of single-qubit and two-qubit basic quantum gates, and there
have been many proposals to implement the Fredkin gate which required at
least six \cite{six}, five \cite{five} and four \cite{four} two-qubit gates.
In particular, the Fredkin gate has been experimentally implemented with
linear optics \cite{16fredk} on qubit systems. However, the multiqubit and
even the high-dimensional quantum gates could be indispensable for the large
scale quantum network and quantum processors. It is usually a significant
challenge to construct such multiqubit or high-dimensional quantum gates
with the increase of the state space. The reason is mainly attributed to not
only the experimental complications added but also the possibility of the
errors caused by decoherence. Thus, it would be desirable to seek for efficient
schemes based on a reliable physical platform to directly construct the
Fredkin gate so as to reduce the operation time and experimental
complications.

In analogy to cavity quantum electrodynamics (QED), the circuit QED system
studying the light-matter interaction,  is a specially suited platform to
realize QCQIP due to its flexibility, scalability, and tunability \cite%
{04Blais,08Clarke,s3,s4}. The strong \cite{04Wallraff,04Chiorescu},
ultrastrong \cite{10Niemczyk,10Forniaz}, and beyond the ultrastrong coupling
regimes \cite{17Yoshihara} with a superconducitng qubit coupled to a
microwave resonator have been experimentally achieved in a series of
experiments, and the strong coupling of an nitrogen-vacancy center ensemble
(NV ensemble) to a superconducting resonator \cite{11nvr,11Kubo} or flux
qubit \cite{11zhu} has been experimentally realized in circuit QED. In
addition, quantum memory is also indispensable in QCQIP such as quantum
repeater and quantum computing \cite{11Simon}. A distinct feature of quantum
memory is that it has the relatively large state space. In recent years,
the solid-state devices (such as NV ensembles and superconducting resonators)
have been considered as the good memory elements in QCQIP \cite%
{11Simon,11Mariantoni}. Up to now, the superconducting resonator lifetimes
between 1 and 10 ms have been reported \cite{13Reagor,16Reagor,16Axline} and
a lifetime of 1 s for an NV ensemble has been experimentally achieved \cite%
{16NC}. These experimental achievements directly lead to the further
breakthrough in QCQIP. In particular, as the important physical resource,
quantum entanglement in the context of circuit QED has attracted a great
many of interest such as the preparation of a variety of entangled states
(e.g., Bell states, NOON states, and entangled coherent states) of two
superconducting resonators \cite%
{08Mariantoni,njpMerkel,14Su,14ma,15Hua,17liu,17Yang,10Strauch,12Strauch,16Sharma,16Zhao}
or NV ensembles \cite{11yangwan,12chen}.  Experimentally, the photon NOON states
of two superconducting resonators have been produced \cite{11Wangh}, and a
two-mode entangled coherent state of microwave fields in two superconducting
resonators has been prepared \cite{16Wangc}. Thus it is natural to consider
how we can construct the Fredkin gate in high-dimensional systems by
utilizing the well developed experimental technology in the superconducting quantum
system.

Here, we propose a method for the \emph{direct} realization of a general hybrid tripartite
Fredkin gate by using superconducting resonators as two quantum
memories coupled to a superconducting flux qubit. This gate can be expressed
as
\begin{eqnarray}  \label{eq1}
U(\gamma|g\rangle+\eta|e\rangle)|\psi\rangle_1|\varphi\rangle_2=\gamma|g%
\rangle
|\varphi\rangle_1|\psi\rangle_2+\eta|e\rangle|\psi\rangle_1|\varphi\rangle_2,
\end{eqnarray}
where $\gamma$ and $\eta$ are the normalized complex numbers,  $%
|\psi\rangle$ and $|\varphi\rangle$ are arbitrary pure states of target
qudits encoded in two quantum memories 1 and 2, and $|g\rangle$ and $%
|e\rangle$ are the states of the control qubit. Eq.~(1)
shows that if and only if the control qubit is in the state $|g\rangle$,
the two target qudits will swap their states, otherwise they remain in their
initial states. Considering the experimental progress made in the NV center, we
also propose an experimental scheme to realize the same aim as above by
utilizing the NV ensembles as the quantum memories.

The two proposals have the following distinct advantages: (i) The Fredkin
gate can be realized by employing a single unitary operation without need of any microwave pulse; (ii) Our method and experimental setup are simple
because only a single qutrit and two target quantum
memories are used; (iii) The experimental scheme is based on the superconducting
resonator or the NV ensemble which has a long coherence time; (iv) Each
controlled target qudit of this gate can be in an arbitrary state (\emph{%
discrete-variable or continuous-variable state}) which can further lead to the
wide applications such as (a) preparing an \emph{arbitrary} entangled state of
two superconducting resonators or NV ensembles, (b) directly measuring the
fidelity between the two quantum memories as well as the entanglement
between them without any information on the initial states required.

This paper is organized as follows. In Sec.~\ref{sec2}, we explicitly show how to
implement the hybrid Fredkin gate of a single superconducting flux qubit
simultaneously controlling two target qudits encoded in two superconducting
coplanar waveguide resonators or NV ensembles. In Sec.~\ref{sec3}, we discuss the
applications and the possible experimental implementation of our proposal
and numerically calculate the operational fidelity for creating NOON states,
entangled coherent states, and entangled cat states of two resonators or NV
ensembles. A concluding summary is given in Sec. \ref{sec4}.

\section{Hybrid Fredkin gate between a single superconducting qubit and two quantum memories} \label{sec2}

\begin{figure}[tbp]
\begin{center}
\includegraphics[bb=66 590 421 754, width=12.5 cm, clip]{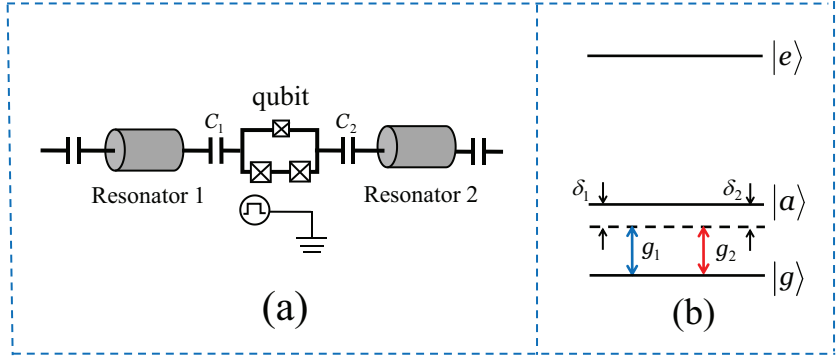} \vspace*{%
-0.08in}
\end{center}
\caption{(a) Setup of two superconducting resonators (i.e., 1 and 2) coupled
to a flux qutrit (coupler) via capacitances $C_1$ and $C_2$. (b) Resonator $1$
($2$) is far-off resonant with $|g\rangle\leftrightarrow|a\rangle$
transition of coupler with coupling strength $g_1$ ($g_2$) and detuning $%
\protect\delta_1$ ($\protect\delta_2$). Here, the detunings $\protect\delta%
_1=\protect\omega_{ag}-\protect\omega_{a_1}$ and $\protect\delta_2=\protect%
\omega_{ag}-\protect\omega_{a_2}$, the $\protect\omega_{ag}$ is the $%
|g\rangle\leftrightarrow|a\rangle $ transition frequency of coupler and the $%
\protect\omega_{a_1}$ ($\protect\omega_{a_2}$) is the frequency of resonator
$1$ ($2$). }
\label{fig:1}
\end{figure}

\textit{Superconducting resonators as quantum memories---}We first consider such
a system that consists of two superconducting microwave coplanar waveguide
resonators coupled to a three-level superconducting flux qutrit (coupler)
[Fig.~1(a)]. As shown in Fig.~1(b), the resonators $1$ and $2$ are
off-resonantly coupled to the $|g\rangle\leftrightarrow|a\rangle $
transition of coupler with the coupling constants $g_1$ and $g_2$, respectively.
In the interaction picture, after making the rotating-wave approximation,
the Hamiltonian of the whole system reads (in units of $\hbar =1$%
)
\begin{eqnarray}  \label{eq2}
H_{I,1}=g_1(e^{i\delta_1 t}a_1\sigma _{ag}^{+}+h.c.)+ g_2(e^{i\delta_2
t}a_2\sigma_{ag}^{+}+h.c.),
\end{eqnarray}
where $a_1$ ($a_2$) is the photon annihilation operator for the resonator $1$
($2$), $\sigma _{ag}^{+}=|a\rangle\langle g|$, $\delta_1=\omega_{ag}-%
\omega_{a_1}$ and $\delta_2=\omega_{ag}-\omega_{a_2}$. Here, $\omega_{ag}$
is the frequency of coupler related to the transition $|g\rangle\leftrightarrow|a\rangle $
and $\omega_{a_1}$ ($\omega_{a_2}$) is the frequency of resonator $1$ ($2$).

Considering the large-detuning conditions $\delta_1\gg g_1$ and $\delta_2\gg g_2
$, the Hamiltonian~(\ref{eq2}) can be written as \cite{s35}
\begin{eqnarray}  \label{eq3}
H_{e} &=&(\frac{g_1^2}{\delta_1} a_1a_1^\dagger+\frac{g_2^2}{\delta_2}a_2
a_2^\dagger)|a\rangle\langle a| -(\frac{g_1^2}{\delta_1}a_1^\dagger a_1 +%
\frac{g_2^2}{\delta_2}a_2^\dagger a_2)|g\rangle\langle g|  \notag \\
&+&\lambda(e^{i\delta^{\prime }t}a_1^\dagger a_2+e^{-i\delta^{\prime }t}a_1
a_2^\dagger)|a\rangle\langle a|-\lambda(e^{-i\delta^{\prime }t}a_1^\dagger
a_2+e^{i\delta^{\prime }t}a_1 a_2^\dagger)|g\rangle\langle g|,
\end{eqnarray}%
where $\lambda=\frac{g_1 g_2}{2}(\frac{1}{\delta_1}+\frac{1}{\delta_2})$ and
$\delta^{\prime }=\delta_2-\delta_1$. The first  line of Eq.~(\ref{eq3}) describes Stark shifts of the level $|a\rangle
$ ($|g\rangle$) of the coupler; which the second line
describes the interaction between the resonators $1$ and $2$ when the
coupler is in the state $|a\rangle$ or $|g\rangle$.

For simplicity, we set
\begin{eqnarray}  \label{eq4}
\delta_1=\delta_2=\delta, g_1=g_2=g,
\end{eqnarray}
and assume that the level $|a\rangle$ of coupler is not occupied. Thus, the
effective Hamiltonian~(\ref{eq3}) is reduced to
\begin{eqnarray}  \label{eq5}
H_{e}&=&H_{0}+H_{i}
\end{eqnarray}
with
\begin{eqnarray}  \label{eq6}
H_{0}&=&-\omega(a_1^\dagger a_1 +a_2^\dagger a_2)|g\rangle\langle g|,  \notag
\\
H_{i}&=&-\lambda(a_1^\dagger a_2+a_1 a_2^\dagger)|g\rangle\langle g|,
\end{eqnarray}
where $\omega=g_1^2/\delta_1=g_2^2/\delta_2$. The Hamiltonian $H_{i}$
describes the interaction between the resonators when the coupler is in the
state $|g\rangle$.

Now we can show that the effective Hamiltonian~(\ref{eq5}) can be used to
construct a hybrid Fredkin gate with the flux qubit simultaneously controlling the
two target qudits encoded by two quantum memories.
We suppose the quantum memory is prepared by a superconducting coplanar waveguide resonator or an
NV ensemble. Let the initial state of the quantum memory $1$ ($2$) be an arbitrary
pure state $|\psi \rangle _{1}=\sum\limits_{n=0}^{\infty }c_{n}|n\rangle _{1}
$ with $|n\rangle _{1}=\frac{(a_{1}^{\dagger })^{n}}{\sqrt{n!}}|0\rangle _{1}
$ ($|\varphi \rangle _{2}=\sum\limits_{m=0}^{\infty }d_{m}|m\rangle _{2}$
with $|m\rangle _{2}=\frac{(a_{2}^{\dagger })^{m}}{\sqrt{m!}}|0\rangle _{2}$%
) and the flux qutrit be an arbitrary superposition state $|\phi \rangle
_{q}=\gamma |g\rangle +\eta |e\rangle $.
Note that only the two levels $|g\rangle$ and $|e\rangle$ of the flux qutrit
are encoded here and will be used as the control qubit in the Fredkin gate.
Here, $c_{n}$ and $d_{m}$ ($\gamma $
and $\eta $) represent arbitrary complex amplitudes satisfying
the normalization condition, $|0\rangle _{1}$ ($|0\rangle _{2}$) denotes
the vacuum state of the quantum memory 1 (2). From Eq.~(\ref{eq6}), one can see that $[H_{0},H_{i}]=0$. Thus,
the time-evolution operator for the Hamiltonian~(\ref{eq5}) can be defined
as $U=e^{-iH_{0}t}\cdot e^{-iH_{i}t}$. With the Hamiltonian~(\ref{eq5}),
the state $|\phi _{q}\rangle |\psi \rangle _{1}$$%
|\varphi \rangle _{2}$ of the qubit-memory system  evolves into
\begin{eqnarray}
U|\phi \rangle _{q}|\psi \rangle _{1}|\varphi \rangle _{2}
&=&e^{-iH_{0}t}e^{-iH_{i}t}(\gamma |g\rangle +\eta |e\rangle )|\psi
\rangle_{1} |\varphi \rangle_{2}   \notag  \label{eq15} \\
&=&e^{i\omega (a_{1}^{\dagger }a_{1}+a_{2}^{\dagger }a_{2})t|g\rangle
\langle g|}e^{i\lambda (a_{1}^{\dagger }a_{2}+a_{1}a_{2}^{\dagger
})t|g\rangle \langle g|}(\gamma |g\rangle +\eta |e\rangle )|\psi \rangle
_{1}|\varphi \rangle _{2}  \notag \\
&=&e^{i\omega (a_{1}^{\dagger }a_{1}+a_{2}^{\dagger }a_{2})t}e^{i\lambda
(a_{1}^{\dagger }a_{2}+a_{1}a_{2}^{\dagger })t}\gamma |g\rangle |\psi
\rangle _{1}|\varphi \rangle _{2}+\eta |e\rangle |\psi \rangle _{1}|\varphi
\rangle _{2}  \notag \\
&=&e^{iH_{0}^{\prime }t}e^{iH_{i}^{\prime }t}\gamma |g\rangle |\psi \rangle
_{1}|\varphi \rangle _{2}+\eta |e\rangle |\psi \rangle _{1}|\varphi \rangle
_{2},
\end{eqnarray}%
where $H_{0}^{\prime }=\omega (a_{1}^{\dagger }a_{1}+a_{2}^{\dagger }a_{2})$%
, $H_{i}^{\prime }=\lambda (a_{1}^{\dagger }a_{2}+a_{1}a_{2}^{\dagger })$,
and $\langle g|e\rangle =0$ has been used. It should be noted that the
Hamiltonian $H_{0}^{\prime }$ and $H_{i}^{\prime }$ are different from $H_{0}
$ and $H_{i}$ because the Hamiltonian $H_{0}$ and $H_{i}$ contain a qubit
operator $|g\rangle \langle g|$.

Making use of the Hamiltonian $H_{i}^{\prime }$, one can obtain the
transformations $e^{-iH_{i}^{\prime }t}a_{1}^{\dagger }e^{iH_{i}^{\prime
}t}=\cos (\lambda t)a_{1}^{\dagger }-i\sin (\lambda t)a_{2}^{\dagger }$ and $%
e^{-iH_{i}^{\prime }t}a_{2}^{\dagger }e^{iH_{i}^{\prime }t}=\cos (\lambda
t)a_{2}^{\dagger }-i\sin (\lambda t)a_{1}^{\dagger }$. For $\lambda t=\pi /2$%
, one has $e^{-iH_{i}^{\prime }t}(a_{1}^{\dagger })^{n}e^{iH_{i}^{\prime
}t}=(-ia_{2}^{\dagger })^{n}$ and $e^{-iH_{i}^{\prime }t}(a_{2}^{\dagger
})^{m}e^{iH_{i}^{\prime }t}=(-ia_{1}^{\dagger })^{m}$. After the evolution
time $t=\pi /(2\lambda )$, the Eq.~(\ref{eq15}) becomes
\begin{eqnarray}
&&U|\phi \rangle _{q}|\psi \rangle _{1}|\varphi \rangle _{2}  \notag
\label{eq16} \\
&=&e^{iH_{0}^{\prime }t}e^{iH_{i}^{\prime }t}\gamma |g\rangle |\psi \rangle
_{1}|\varphi \rangle _{2}+\eta |e\rangle |\psi \rangle _{1}|\varphi \rangle
_{2}  \notag \\
&=&e^{iH_{0}^{\prime }t}\gamma |g\rangle \sum\limits_{n=0}^{\infty
}\sum\limits_{m=0}^{\infty }\frac{c_{n}}{\sqrt{n!}}\frac{d_{m}}{\sqrt{m!}}%
e^{iH_{i}^{\prime }t}(a_{1}^{\dagger })^{n}(a_{2}^{\dagger })^{m}|0\rangle
_{1}|0\rangle _{2}+\eta |e\rangle |\psi \rangle _{1}|\varphi \rangle _{2}
\notag \\
&=&e^{iH_{0}^{\prime }t}\gamma |g\rangle \sum\limits_{n=0}^{\infty
}\sum\limits_{m=0}^{\infty }\frac{c_{n}}{\sqrt{n!}}\frac{d_{m}}{\sqrt{m!}}%
\left[ e^{iH_{i}^{\prime }t}(a_{1}^{\dagger })^{n}e^{-iH_{i}^{\prime }t}%
\right] \left[ e^{iH_{i}^{\prime }t}(a_{2}^{\dagger })^{m}e^{-iH_{i}^{\prime
}t}\right] e^{iH_{i}^{\prime }t}|0\rangle _{1}|0\rangle _{2}
 \notag \\
&+&\eta |e\rangle
|\psi \rangle _{1}|\varphi \rangle _{2}  \notag \\
&=&e^{iH_{0}^{\prime }t}\gamma |g\rangle \sum\limits_{m=0}^{\infty }\frac{d_{m}}{\sqrt{m!}}(-ia_{1}^{\dagger
})^{m}|0\rangle _{1} \sum\limits_{n=0}^{\infty }\frac{%
c_{n}}{\sqrt{n!}}(-ia_{2}^{\dagger })^{n}|0\rangle
_{2}+\eta |e\rangle |\psi \rangle _{1}|\varphi \rangle _{2}
\notag \\
&=&\gamma |g\rangle |\varphi \rangle _{1}|\psi \rangle _{2}+\eta |e\rangle
|\psi \rangle _{1}|\varphi \rangle _{2},
\end{eqnarray}%
where we have used $a_{1}^{\dagger }a_{1}|n\rangle =n$ ($a_{2}^{\dagger
}a_{2}|m\rangle =m$), $a_{1}|0\rangle _{1}=0$ ($a_{2}|0\rangle _{2}=0$), and
$(-i)^{n}=e^{in\pi (2k-1/2)}$ ($(-i)^{m}=e^{im\pi (2k-1/2)})$ with $k$ an integer.
One finds that the phase factors
$(-i)^{n}$ and $(-i)^{m}$ in Eq.~(\ref{eq16}) have been completely dropped for the evolution time $t=\pi /(2\lambda )$.
Eq.~(\ref{eq16}) is a hybrid Fredkin gate which
implies that iff the control qubit is in the state $|g\rangle $, the two
target qudits swap their states but nothing otherwise.

We would like to mention that various schemes for achieving the three-qubit Fredkin
gates have been previously suggested in \cite%
{gate1,gate2,gate3,gate4,gate5,gate6,gate7}, in which the two target qubits
are encoded in two natural/artificial atoms \cite%
{gate1,gate2,gate3,gate4,gate5,gate6} or photons \cite{gate7}. Unlike
the previous proposals, the target
qubits of our gate are encoded in two superconducting resonators or NV
ensembles which have the long coherence times. In contrast to~\cite{gate6} where
 the target qubits are encoded by the discrete-variable state, in our
proposal the target qubits also can be encoded by the continuous-variable state. In addition,
the proposals \cite{gate1,gate2,gate3,gate4,gate5,gate6,gate7} require
several operational steps and the microwave pulses, however, our proposal is
much improved because our gate can be realized only using a single-step
operation and no microwave pulses are needed.

\textit{NV ensembles as quantum memories---} Now we would like to consider
another system composing of a superconducting flux qubit coupled to two NV
ensembles to realize the above Fredkin gate.
\begin{figure}[tbp]
\begin{center}
\includegraphics[bb=79 585 526 745, width=12 cm, clip]{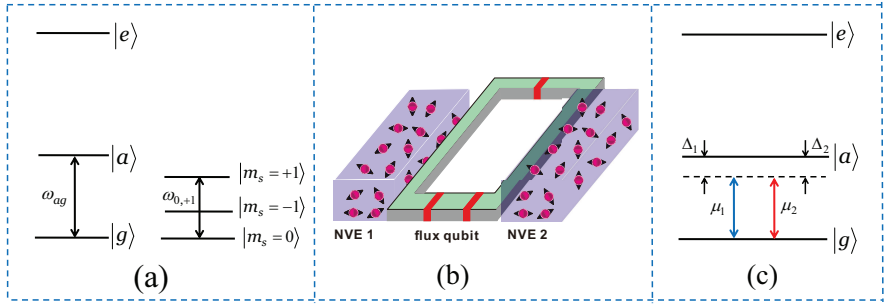} \vspace*{%
-0.08in}
\end{center}
\caption{ (a) Energy level diagram of the flux and an NV center. By applying
an external magnetic field along the crystalline axis of the NV center, an
additional Zeeman splitting between $|m_{s}=\pm 1\rangle $ sublevels occurs.
Here, $\protect\omega_{ag}$ is the $|g\rangle\leftrightarrow|a\rangle $
transition frequency of flux qubit and $\protect\omega_{0,+1}$ is the energy
gap between the $|m_s=0\rangle $ and $|m_s=+1\rangle $ levels of the NV
center. (b) Illustration of the hybrid system consisted of a flux qubit and
two NV ensembles. (c) The NV ensemble $1$ ($2)$ is far-off resonant with $%
|g\rangle\leftrightarrow|a\rangle$ transition of qubit with the coupling
strength $\protect\mu_1$ ($\protect\mu_2$) and the detuning $\Delta_1$ ($\Delta_2
$). }
\label{fig:2}
\end{figure}
The energy-level of an NV center consists of a ground state $^{3}A$, an
excited state $^{3}E$ and a metastable state $^{1}A$. Both $^{3}A$ and $^{3}E
$ are spin triplet states while the metastable $^{1}A$ is a spin singlet
state~\cite{96Lenef,06Manson}. The NV center has an electronic spin triplet
ground state with zero-field splitting $D_{gs}/\left( 2\pi \right)
\approx2.878$ GHz between the $|m_{s}=0\rangle $ and $|m_{s}=\pm 1\rangle $
levels. By applying an external magnetic field along the crystalline axis of
the NV center~\cite{06Neumann1,13xiang}, an additional Zeeman splitting
between $|m_{s}=\pm 1\rangle $ sublevels occurs [Fig.~2(a)].

We first consider a system consisting of a superconducting flux qubit coupled
to an NV ensemble. The NV center is usually regarded as a spin while an NV
ensemble is generally considered as a spin ensemble. We choose the $%
|g\rangle\leftrightarrow|a\rangle $ transition of qubit is coupled to the
transition between the ground level $|m_{s}=0\rangle $ and the excited level
$|m_{s}=+1\rangle $ of the spins in the ensemble, but decoupled from the
transition between the two levels $|m_{s}=0\rangle $ and $|m_{s}=-1\rangle$.
In the interaction picture, after making the rotating-wave approximation,
the Hamiltonian of the flux qubit and the NV ensemble system is
\begin{equation}  \label{eq7}
H_{FN}=\sum\limits_{k=1}^{N}\mu_{k}(\sigma^{+}_{ag}\tau _{k}^{-}e^{i\Delta
t}+\sigma^{-}_{ag}\tau _{k}^{+}e^{-i\Delta t}),
\end{equation}%
where $\Delta =\omega _{ag}-\omega _{0,+1}$, $\tau _{k}^{-}=|m_{s}=0\rangle
_{k}\langle m_{s}=+1|$ and $\tau _{k}^{+}=|m_{s}=+1\rangle _{k}\langle
m_{s}=0|$ are the lowering and raising operators of the $k$th spin, and $%
\mu_{k}$ is the coupling constant between the $k$th spin and the $%
|g\rangle\leftrightarrow|a\rangle $ transition of qubit. Here, $\omega
_{0,+1}$ is the transition frequency between the two levels $|m_{s}=0\rangle
$ and $|m_{s}=+1\rangle$. We then introduce a collective operator
\begin{equation}  \label{eq8}
b^{\dagger }=\left( \frac{1}{\sqrt{N}}\right) \left( \frac{1}{\overline{\mu}}%
\right) \sum\limits_{k=1}^{N}\mu_{k}\tau _{k}^{+},
\end{equation}%
where $\overline{\mu}^{2}=\sum\limits_{k=1}^{N}|\mu_{k}|^{2}/N$ with $%
\overline{\mu} $ the root mean square of the individual couplings.

Under the conditions of the large $N$ and the low excitations, $b^{\dagger }$
behaves as a bosonic operator and the spin ensemble behaves as a bosonic
mode. Thus, one has $[b,b^{\dagger }]\approx 1,$ and $b^{\dagger }b|n\rangle
_{b}=n|n\rangle _{b}$~\cite{13xiang,12Hmmer}, where $|n\rangle _{b}=\frac{1}{%
\sqrt{n!}}(b^{\dagger })^{n}|0\rangle _{b}$ with $|0\rangle
_{b}=|m_{s}=0\rangle _{1}|m_{s}=0\rangle _{2}\cdots |m_{s}=0\rangle _{N}$.
Accordingly, one has the frequency of the bosonic mode $\omega _{b}=\omega
_{0,+1}$.Therefore, the Hamiltonian~(\ref{eq7}) can be further rewritten as
\begin{equation}  \label{eq9}
H_{FN}=\mu(e^{i\Delta t}b\sigma^{+}_{ag}+e^{-i\Delta t}b^{\dagger
}\sigma^{-}_{ag}),
\end{equation}%
with $\mu=\sqrt{N}\overline{\mu}$. We then consider a system consisting of a
flux qubit coupled to two NV ensembles [Fig.~2(b)]. As depicted in
Fig.~2(c), NV ensembles 1 and 2 are off-resonantly coupled to the $%
|g\rangle\leftrightarrow|a\rangle $ transition of qubit with coupling
constants $\mu_1$ and $\mu_2$, respectively. Based on Eq.~(\ref{eq9}), the
Hamiltonian of the whole system is
\begin{eqnarray}  \label{eq10}
H_{I,1}=\mu_1(e^{i\Delta_1 t}b_1\sigma _{ag}^{+}+h.c.)+ \mu_2(e^{i\Delta_2
t}b_2\sigma_{ag}^{+}+h.c.),
\end{eqnarray}
where $b_1$ and $b_2$ are the corresponding annihilation operators for the
NV ensembles 1 and 2, $\Delta_1=\omega_{ag}-\omega_{b_1}$ and $%
\Delta_2=\omega_{ag}-\omega_{b_2}$. Here, $\omega_{ag}$ is the $%
|g\rangle\leftrightarrow|a\rangle $ transition frequency of qubit and $%
\omega_{b_1}$ ($\omega_{b_2}$) is the frequency of NV ensemble $1$ ($2$).

Let's assume the large-detuning conditions $\Delta_1\gg \mu_1$ and $\Delta_2\gg\mu_2
$ with  (i) $\Delta_1=\Delta_2,$ $\mu_1=\mu_2,$ (ii) the level $%
|a\rangle$ of coupler qubit not occupied. One can find that the final
effective Hamiltonian [for details, see Eqs.~(\ref{eq3})-(\ref{eq5})] can be
given by
\begin{eqnarray}  \label{eq11}
H_{e}&=&H_{0}+H_{i}
\end{eqnarray}
with
\begin{eqnarray}  \label{eq12}
H_{0}&=&-\omega(b_1^\dagger b_1 +b_2^\dagger b_2)|g\rangle\langle g|,  \notag
\\
H_{i}&=&-\lambda(b_1^\dagger b_2+b_1 b_2^\dagger)|g\rangle\langle g|,
\end{eqnarray}
where $\omega=\mu_1^2/\Delta_1=\mu_2^2/\Delta_2$ and $\lambda=\frac{\mu_1
\mu_2}{2}(\frac{1}{\Delta_1}+\frac{1}{\Delta_2})$. When the qubit is in the
state $|g\rangle$, the Hamiltonian $H_{i}$ describes the interaction between
the NV ensembles. It is obvious that the Hamiltonian Eq. (\ref{eq11}) has
the same form as Eq. (\ref{eq5}), so the above demonstrated Fredkin gate can
also be prepared with the NV ensembles as memories.

\section{Applications and possible experimental implementation}
\label{sec3}

The Fredkin gate has played a vital role and has many useful applications in QCQIP, such as error correction \cite%
{correction1,correction2}, quantum cloning \cite{cloning}, quantum
fingerprinting \cite{fingerprinting1,fingerprinting2}, and quantum digital
signatures \cite{digital}. In the following subsections, we will discuss some additional applications
and the possible experimental implementation of our proposed Fredkin gate.

\textit{Preparation of entanglement---}Entanglement as a physical phenomenon, is one of the most fundamental features of quantum mechanics \cite{s1}.
Furthermore, entanglement is also an important physical resource to achieve many quantum information processing and communication tasks. For instance, the NOON
states play the central role in quantum metrology \cite{metrology1}, quantum
optical lithography \cite{lithography}, and precision measurement \cite%
{10measurements}; the entangled coherent states can  serve as an
important resource for quantum networks \cite{network}, quantum
teleportation \cite{teleportation}, quantum cryptography \cite{cryptography}%
, and quantum metrology \cite{metrology2}. Thus, the generation of entangled
states is one of the key goals of QCQIP. Over the past two
decades, many experiments have been reported for the generation of
multiple-particle entangled states of photons \cite{11guo,12pan}, ions \cite%
{00ion,11ion}, natural atoms \cite{00atom,01atom}, NV centers \cite%
{08Neumann,13Bernien}, and superconducting qubits \cite%
{10three1,10three2,10bit}.

Our Fredkin gate can be used as an efficient quantum generator to produce
the entanglement between two resonators or NV ensembles with the arbitrary
(discrete variable or continuous variable) initial state. For example, we
apply a microwave pulse to control qubit such that the pulse is resonant
with the $|g\rangle \leftrightarrow |e\rangle $ transition of the control
qubit. Making the rotating-wave approximation, the Hamiltonian in the interaction picture is written as $%
H_{I_{,}2}=\Omega (e^{i\theta }|g\rangle \langle e|+h.c.),$ where $\Omega $
and $\theta $ are the Rabi frequency and the initial phase of the pulse. We
choose $t^{\prime }=\pi /\left( 4\Omega \right) $ and $\theta =-\pi /2$ to
pump the state $|e\rangle $ to $(|e\rangle -|g\rangle )/\sqrt{2}$ and $%
|g\rangle $ to $(|e\rangle +|g\rangle )/\sqrt{2}$. Accordingly, the state~(%
\ref{eq16}) changes to $\frac{1}{\sqrt{2}}(|\psi _{+}\rangle |e\rangle
+|\psi _{-}\rangle |g\rangle )$, where $|\psi _{\pm }\rangle $ are the
entangled states of two quantum memories, given by $|\psi _{\pm }\rangle
=\gamma |\varphi \rangle _{1}|\psi \rangle _{2}\pm \eta |\psi \rangle
_{1}|\varphi \rangle _{2}.$ Now if a von Neumann measurement is performed on
flux qubit along a measurement basis \{$\left\vert g\right\rangle
,\left\vert e\right\rangle $\}, one can see that the entangled state of two
quantum memories is prepared in the $|\psi _{+}\rangle $ or $|\psi
_{-}\rangle $. It should be noted here that  $|\psi \rangle $ and $%
|\varphi \rangle $ are arbitrary nonsymmetric states.

The previous schemes for synthesizing an arbitrary two-resonator entangled
state through two three-level superconducting qutrits coupled to three
resonators assisted by a sequence of microwave pulses applied to the two
qubits \cite{njpMerkel}, or a four-level superconducting qudit \cite%
{17Yang} coupled to two resonators and driven by a microwave pulse. Compared
with \cite{njpMerkel,17Yang}, our experimental setup is
greatly simplified and the experimental difficulty is reduced because only a single three-level qubit is employed and no microwave
pulse is used. In addition, our proposal is based on a first-order large detuning,
while the ~\cite{17Yang} was based on a second-order large detuning.
This makes our operation faster than the one proposed by \cite{17Yang}. In
recent years, several state-synthesis algorithms have been proposed to
generate entangled states of two superconducting resonators \cite%
{10Strauch,12Strauch,16Sharma,16Zhao}. These methods \cite%
{10Strauch,12Strauch,16Sharma,16Zhao} depend on the maximum photon number
and the number of operational steps required. While our proposal requires
only a single unitary operation, which can significantly reduce the number
of steps and the preparation time.

\textit{Measuring the fidelity and entanglement between the two quantum
memories---} Determining the overlap (or fidelity) of one quantum state with respect to another or the entanglement of a pure state based on the controlled-swap gate is a well-known method usually employed in qubit systems \cite{Filip,Brennen,Horo,YCS}.
Here as an alternative demonstration in qudit systems, we will show that the above experimental scheme can also be used to determine the fidelity and entanglement between the two quantum
memories. Let's consider such a scenario where the two quantum memories
1 and 2 are in a product state $\left\vert \psi \right\rangle
_{1}\left\vert \varphi \right\rangle _{2}$ and one wants to know whether $%
\left\vert \psi \right\rangle _{1}$ and $\left\vert \varphi \right\rangle
_{2}$ are the same or not or to what degree they are different. To achieve
this goal, one can first prepare the flux qubit in the state $|\phi \rangle
_{q}=\gamma |g\rangle +\eta |e\rangle $ with the known parameter $\gamma $
and $\eta .$ Typically, we can set $\gamma =\eta =\frac{1}{\sqrt{2}}$. Let
the  tripartite quantum state $|\phi \rangle _{q}\left\vert \psi
\right\rangle _{1}\left\vert \varphi \right\rangle _{2}$ undergo the above
preparation process of the entanglement, one can find that, before the final
von Neumann measurement on the flux qubit,  $|\phi \rangle _{q}\left\vert
\psi \right\rangle _{1}\left\vert \varphi \right\rangle _{2}$ will arrive at
\begin{eqnarray}
|\phi \rangle _{q}\left\vert \psi \right\rangle _{1}\left\vert \varphi
\right\rangle _{2} &\longrightarrow &\frac{1}{\sqrt{2}}|g\rangle \left(
\gamma \left\vert \varphi \right\rangle _{1}\left\vert \psi \right\rangle
_{2}-\eta \left\vert \psi \right\rangle _{1}\left\vert \varphi \right\rangle
_{2}\right)   \notag \\
&&+\frac{1}{\sqrt{2}}|e\rangle \left( \gamma \left\vert \varphi \right\rangle
_{1}\left\vert \psi \right\rangle _{2}+\eta \left\vert \psi
\right\rangle _{1}\left\vert \varphi \right\rangle _{2}\right) \text{.}
\end{eqnarray}%
Now let's preform the von Neumann measurement on the flux qubit with respect
to the basis $\left\{ |g\rangle ,|e\rangle \right\} $, one can immediately
find that the probability collapsing on the state $|g\rangle $ is given by%
\begin{equation}
p_{g}=\frac{1}{2}[1-\left( \gamma ^{\ast }\eta +\gamma \eta ^{\ast }\right) F^{2}
\label{fi1}]
\end{equation}%
with $F=\left\vert \left\langle \varphi \right. \left\vert \psi
\right\rangle \right\vert $ denoting the fidelity between the two states.
Thus for any nonvanishing $\gamma ^{\ast }\eta +\gamma \eta ^{\ast }$ known
beforehand, one can easily find the fidelity of the two states of the two
memories by the measurement probability.

In fact, the above procedure can also be used to determine the entanglement
that we have prepared in the above section. As usual, to prepare
entanglement, one has to prepare the initial state like $|\phi \rangle _{q}\left\vert
\psi \right\rangle _{1}\left\vert \varphi \right\rangle _{2}$%
. Since all the information of the initial state is known, one can calculate
the entanglement by some proper entanglement measure such as the concurrence
\cite{wootters} defined by%
\begin{equation}
C(|\psi _{\pm }\rangle )=\sqrt{2[1-\mathrm{Tr}(\rho _{r}^{2})]}
\label{conc}
\end{equation}%
with $\rho _{r}$ representing the reduced density matrix of the state $|\psi
_{\pm }\rangle $. Here, we would like to emphasize that actually this
concurrence can be measured directly based on the above experimental
procedure. In particular, if we \textit{don't know the initial states} in
which the two memories are, one can find that the entanglement prepared
through the above procedure can also be directly measured. To see this,
let's substitute the state $|\psi _{\pm }\rangle $ into Eq. (\ref{conc}).
One can immediately find that
\begin{equation}
C(|\psi _{\pm }\rangle )=\sqrt{2-\frac{1}{2}\{\left|\gamma|^4+|\eta|^4+2\left[ 2|\gamma|^2 |\eta|^2
\pm(\gamma\eta^*+\gamma^*\eta)\right]F^2+(\gamma^2\eta^{*2}+\gamma^{*2}\eta^2)F^4\right\}}.  \label{conc2}
\end{equation}%
It is obvious that $F^{2}$ in Eq. (\ref{conc2}) can be measured based on the
measurement of the fidelity given by Eq. (\ref{fi1}), since  $\gamma $ and $%
\eta \ $\ are known beforehand.

\textit{Experimental feasibility---}Recent experiments have achieved
arbitrary control of a single superconducting resonator in circuit QED~\cite%
{08Hofheinz1,09Hofheinz2,13Vlastakis}. For instance,~\cite{08Hofheinz1}
experimentally generated Fock states in a superconducting resonator by
coupling a tunable superconducting qubit,~\cite{09Hofheinz2}
demonstrated the preparation of arbitrary quantum states of a single
resonator,~\cite{13Vlastakis} experimentally created a superposition of
coherent states (i.e., a cat state) of a resonator, and~\cite{17Heeres}
experimentally implemented a universal set of gates on a qubit encoded into
a resonator using the cat-code. Here we will take the preparation of
entanglement as examples to give a necessary analysis on the feasibility of
our scheme. For an experimental implementation, we consider the qubit-memory
system initially state is
\begin{eqnarray}
&&\text{(i)}~|\phi \rangle_{q} |N\rangle _{1}|0\rangle _{2}, \\
&&\text{(ii)}~|\phi \rangle_{q} |\alpha \rangle _{1}|-\beta \rangle _{2}, \\
&&\text{(iii)}~|\phi \rangle_{q} |\psi _{e}\rangle _{1}|\varphi _{o}\rangle
_{2},
\end{eqnarray}%
where $|\phi \rangle_{q} =\frac{1}{\sqrt{2}}(|g\rangle +|e\rangle )$ (i.e.,$%
\gamma =\eta =\frac{1}{\sqrt{2}}$), $|\psi _{e}\rangle _{1}=N_{e}(|\alpha
\rangle _{1}+|-\alpha \rangle _{1})$, and $|\varphi _{o}\rangle
_{2}=N_{o}(|\beta \rangle _{2}-|-\beta \rangle _{2})$. Here, $|\alpha
\rangle _{1}$ and $|\beta \rangle _{2}$ ($|-\alpha \rangle _{1}$ and $%
|-\beta \rangle _{2}$) are coherent states of two memories 1 and 2, $N_{e}=%
\frac{1}{\sqrt{2}}[1+\exp (-2\alpha ^{2})]^{-1/2}$ and $N_{o}=\frac{1}{\sqrt{%
2}}[1-\exp (-2\beta ^{2})]^{-1/2}$ are normalization factors.

For a flux qubit, the transition frequency between two neighbor levels is typically 1 to 20~GHz.
Thus, the $|a\rangle\leftrightarrow|e\rangle $ transition frequency of flux
qubit can be highly detuned from the resonator frequency. Accordingly,
the coupling effect of the resonator with the $|a\rangle\leftrightarrow|e\rangle $ transition
is negligibly small, which is thus not considered in the numerical simulation for simplicity.

When the dissipation and the dephasing are included, the dynamics of the
lossy system is determined by the following Lindblad master equation
\begin{eqnarray}  \label{eq24}
\frac{d\rho }{dt} &=&-i[{H}_{I,k},\rho ] +\sum_{i=1,2}\kappa _{i} \mathcal{L}%
[a_i]  \notag \\
&+&\gamma _{ag}\mathcal{L}[ \sigma _{ag}^{-}] +\gamma _{ea}\mathcal{L}[
\sigma _{ea}^{-}]+\gamma _{eg}\mathcal{L}[ \sigma _{eg}^{-}]  \notag \\
&+&\sum_{j=a,e}\left\{ \gamma _{\varphi j}\left( \sigma _{jj}\rho \sigma
_{jj}-\sigma _{jj}\rho /2-\rho \sigma _{jj}/2\right) \right\},
\end{eqnarray}
where ${H_{I,k}}$ is either $H_{I,1}$ or $H_{I,2}$, $i$ represents quantum
memory (i.e., resonator or NV ensemble) $i$ ($i=1,2)$, $\sigma
_{ag}^{-}=\left\vert g\right\rangle\left\langle a\right\vert$, $\sigma
_{ea}^{-}=\left\vert a\right\rangle \left\langle e\right\vert$, $\sigma
_{eg}^{-}=\left\vert g\right\rangle \left\langle e\right\vert , \sigma
_{jj}=\left\vert j\right\rangle\left\langle j\right\vert (j=a,e);$ and $%
\mathcal{L}\left[ \Lambda \right] =\Lambda \rho \Lambda ^{+}-\Lambda
^{+}\Lambda \rho /2-\rho \Lambda ^{+}\Lambda /2,$ with $\Lambda=a_i,\sigma
_{ag}^{-},\sigma _{ea}^{-},\sigma _{eg}^{-}.$ Here, $\kappa_{i}$ is the
decay rates of quantum memory $i$ ($i=1,2$). In addition, $\gamma _{ag}$ is
the energy relaxation rate of the level $\left\vert a\right\rangle $ of
qubit , $\gamma _{ea}$ ($\gamma _{eg}$) is the energy relaxation rate of the
level $\left\vert e\right\rangle $ of qubit for the decay path $\left\vert
e\right\rangle \rightarrow \left\vert a\right\rangle $ ($\left\vert
g\right\rangle $), and $\gamma _{\varphi j}$ is the dephasing rate of the
level $\left\vert j\right\rangle $ of qubit ($j=a,e$).

The fidelity of the operation takes the form $\mathcal{F}=\sqrt{\left\langle
\psi _{\mathrm{id}}\right\vert \rho \left\vert \psi _{\mathrm{id}%
}\right\rangle},$ where $\left\vert \psi _{\mathrm{id}}\right\rangle $ is
the output state of an ideal system (i.e., without dissipation and
dephasing) and $\rho$ is the final density operator of the system when the
operation is performed in a realistic situation. Here, the output state $%
\left\vert \psi _{\mathrm{id}}\right\rangle $ is given by $\frac{1}{\sqrt{2}}%
(|\psi_{+}\rangle|e\rangle+|\psi_{-}\rangle|g\rangle)$. Based on
Eqs.~(19)-(21), the ideal entangled state of two quantum memories is
\begin{eqnarray}
&&\text{(i)}~|N\rangle_1|0\rangle_2\pm|0\rangle_1|N\rangle_2, \\
&&\text{(ii)}~|\alpha\rangle_1|-\beta\rangle_2\pm|-\beta\rangle_1|\alpha%
\rangle_2, \\
&&\text{(iii)}~|\psi_e\rangle_1|\varphi_o\rangle_2\pm|\varphi_o\rangle_1|%
\psi_e\rangle_2.
\end{eqnarray}

By solving the master equation~(\ref{eq24}), the fidelity of the entangled
state generation can be calculated for Eqs.~(23)-(25), respectively. We now
numerically calculate the fidelity for the operation. We choose
resonator-qubit or NV ensemble-qubit coupling constants $g/2 \pi= 70$ MHz.
The values of $g$ here are available in experiments because the
resonator-qubit and the NV ensemble-qubit coupling strengths are
approximately $636$ MHz \cite{10Niemczyk} and $70$ MHz \cite{11zhu} reported in experiments. In addition, we set $k$=1, $N=5$, $%
\alpha=\beta=1.1$, $\Omega/2\pi=100$ MHz (attainable in experiments \cite%
{09pluse,14pluse}), $\gamma^{-1} _{j,\varphi e}=\gamma ^{-1}_{j,\varphi
f}=2~\mu s$, and $\gamma^{-1} _{eg}=\gamma ^{-1}_{fe}=\gamma
^{-1}_{fg}=5~\mu s$. Here we consider a rather conservative case for the
decoherence times of flux qubit~\cite{Pop,s13}. Furthermore, one sets the
lifetimes of two quantum memories (i.e., resonators or NV ensembles) $%
\kappa_1^{-1}=\kappa_2^{-1}=5~\mu s$, which are also conservative estimate
compared with those reported in experiments \cite{s40,s41,16NC}.

\begin{figure}[tbp]
\begin{center}
\includegraphics[bb=0 620 595 756, width=11 cm, clip]{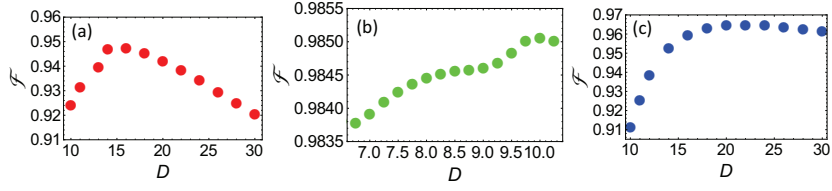} \vspace*{%
-0.08in}
\end{center}
\caption{Fidelity $\mathcal{F}$ versus $D=\protect\delta/g$. (a) Fidelity
versus $D$ for NOON states. (b) Fidelity versus $D$ for entangled coherent
states. (c) Fidelity versus $D$ for entangled cat states. The parameters
used in the numerical simulation are referred in the text.}
\label{fig:3}
\end{figure}

Figure~3 shows the fidelity versus $D=\delta/g$, which are plotted for the
(i) NOON states [Eq.~(23)], (ii) entangled coherent states [Eq.~(24)], (iii)
entangled cat states [Eq.~(25)], respectively. From Figs.~3(a)-3(c), one can
obtain that a high fidelity $96.0\%$, $98.5\%$ and $96.5\%$ for $D=16$, 10
and 22, respectively. In the following analysis, we will choose $D=16$, 10
and 22 for the cases of (i) (ii) and (iii), respectively. For $D=16$, 10 and
22, one has the resonator-qubit or NV ensemble-qubit frequency detunings $%
\delta/2\pi=1.12$ GHz, 0.7 GHz and 1.54 GHz. The numerical simulation shows
that the high-fidelity generation of above entangled states of two quantum
memories is feasible with the state-of-the-art circuit QED technology.
\begin{figure}[tbp]
\begin{center}
\includegraphics[bb=2 283 600 779, width=10 cm, clip]{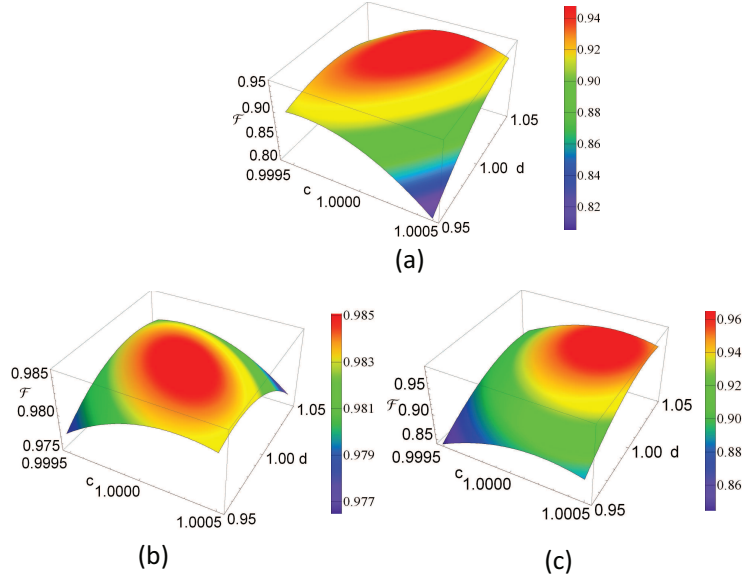} \vspace*{%
-0.08in}
\end{center}
\caption{Fidelity versus $c$ and $d$, which are plotted by choosing (a)~$D=16
$,~(b)~$D=10$, and (c)~$D=22$, respectively.}
\label{fig:4}
\end{figure}

In a realistic situation, the inhomogeneous broadening of quantum memories
may induce the inhomogeneous memory-qubit coupling and unequal memory-qubit
frequency detuning. Thus, we numerically calculate the fidelity by setting $%
\delta_1/2\pi=\delta$, $\delta_2/2\pi=c~\delta$, $g_1=g$, and $g_2=d~g$,
with $c\in[0.9995,1.0005]$ and $d\in[0.95,1.05]$. The other parameters used
in the numerical simulation for Fig.~4 are the same as those used in Fig.~3.
Figures 4(a)-4(c) display the fidelity versus $c$ and $d$, which are plotted
by choosing $D=16$, 10 and 22, respectively. Figure~4(a) shows that for $c\in%
[0.9995,1.0003]$ and $d\in[0.98,1.05]$, the fidelity can be greater than $%
90\%$. As illustrated in Fig.~4(b), the effect of the inhomogeneous
broadening on the fidelity is very small with $c\in[0.9995,1.0005]$ and $d\in%
[0.95,1.05]$. Figure~4(c) displays that the the fidelity is greater than $%
92\%$ for $c\in[0.9998,1.0005]$ and $d\in[0.97,1.05]$. From Fig.~4, one can
see that the high-fidelity generation of entangled states of two quantum
memories can be achieved for small errors in memory-qubit coupling and
detuning.

\section{Conclusion}
\label{sec4}

We have proposed a method to implementing a hybrid Fredkin gate between a
single superconducting flux qubit and two resonators or NV ensembles in any
discrete-variable or continuous-variable states. Due to the usage of only one single
three-level qubit, the experimental setup is simplified and the experimental
difficulty is greatly reduced. In addition, the protocol requires only a
single unitary operation, thus the operation procedure is greatly
simplified. The proposal can also be applied to other kinds of
superconducting qubits (e.g., superconducting charge qubits, transmon
qubits, Xmon qubits, phase qubits) coupled to two 1D resonators or two 3D
cavities. Furthermore, our scheme can be used to generate arbitrary
entangled states of two quantum memories, such as NOON states, entangled
coherent states, and entangled cat states. It is also shown that our scheme
can be used to measure the fidelity and entanglement between the two
memories. Numerical simulation shows that these entangled states can be
high-fidelity created with circuit QED of the existing technology.

Finally, we would like to emphasize that our scheme can also be used to realize a controlled-shift operation with a qudit as a control state. If the control qudit includes two particular energy levels $\left\vert up\right\rangle$ and $\left\vert down\right\rangle$ with the transition frequency much less than the other pairs of energy levels, the two target memories will swap their states conditioned on the control state $\left\vert down\right\rangle$. Similarly, if the highest energy level is much farther off resonant with the other energy levels, one could use the other energy levels as a whole to control the swapping of the states of the two memories. All the realizations need us to carefully choose the various parameters covered in the system. It should be noted that the introduction of the multiple energy levels usually lead to the complex undesired couplings which could be small to some good approximation, but could be a little larger than the case of less energy levels under the same condition. All the above can be easily demonstrated following the similar procedures as our main text. Our
finding provides a new way for realizing the Fredkin gate or quantum
entanglement between two quantum memories, which may have many potential applications in
quantum information processing based on circuit QED.

\section*{Funding}

National Natural Science Foundation
of China (NSFC) (11775040, 11375036); Xinghai Scholar Cultivation
Plan.

\end{document}